\documentclass{amsart}

\usepackage{graphicx}
\usepackage{amsmath,amssymb} 

\newcommand{\be}{\begin{equation}} 
\newcommand{\ee}{\end{equation}} 
\newcommand{\bdis}{\begin{displaymath}} 
\newcommand{\edis}{\end{displaymath}} 
\newcommand{\mbb}{\mathbb} 
\newcommand{\mcal}{\mathcal}

\DeclareMathOperator{\tr}{Tr} 
\DeclareMathOperator{\re}{Re} 
\DeclareMathOperator{\im}{Im}

\newtheorem{theorem}{Theorem}[section]
\theoremstyle{lemma}
\newtheorem{lemma}{Lemma}

\theoremstyle{definition}

\newtheorem{example}{Example}

\theoremstyle{remark}

\numberwithin{equation}{section}

\begin{document}

\title[Aggregate models of liquidity-profit dynamics]{Aggregate models of liquidity-profit dynamics} 

\author{Michal Denetrian} 
\address{Faculty of Mathematics, Physics and Informatics, Comenius University in Bratislava, SLOVAKIA} 
\email{michal.demetrian@uniba.sk} 
\author{Rudolf Zimka} 
\address{Matej Bel University, Bansk\' a Bystrica, SLOVAKIA} 
\email{rudolf.zimka@umb.sk}

\begin{abstract}
We discuss the problem of limit cycles in an aggregate-type models of liquidity growth proposed and studied both theoretically and numerically by W. Semmler and M. Sieveking in \cite{SemmlerSieveking}. Their model is locally of predator-prey type with cannibalism in predator (profit) and logistic bound on prey's population (liquidity). We modify the model to include the weak Allee effect in liquidity and to study the local bifurcations of equilibria and limit cycles by means of the Hopf-Andronov theory. The main motivation for this study follows the economic theory of the so-called corridors of stability introduced by Leijonhuvud in \cite{Leijonh}. Similar importance of these effects is, however, found also in the originally developed dynamical models in biology, especially in mathematical ecology. 
\end{abstract}

\maketitle


\section{Introduction and preliminary results} 

\subsection{Introduction} 

Semmler and Sieveking \cite{SemmlerSieveking} have proposed a dynamical model for the interaction of liquidity and profit in some aggregate sense in accordance with previously published papers considering economic dynamics containing profit vs liquidity interaction, see \cite{Foley}. Semmler and Sieveking were especially interested in the problem of limit cycles of their model, and more specifically in the so-called \emph{corridors of stability}, which were be introduced in the economic theory by Leijonhuvud in \cite{Leijonh}, and subsequently discussed in many papers in the field of economic dynamics. 

In particular, the model of Semmler and Sieveking is based on a modification of the Lotka-Volterra type dynamics with logistic bound on the prey's population and cannibalism in predators. This model, however, has, if any, locally stable positive-quadrant equilibrium, that is also a global attractor in the positive quadrant. To overcome this issue of the model, Semmler and Sieveking proposed a mechanism of discrimination of firms asking for bank loans if these firms are not in somehow good conditions by means of their liquidity (and profit). The equations in questions read 
\be \label{SSE} 
\begin{split}
& \dot{x}=x\left(1-\frac{x}{K}\right)-xy-x\mcal{L}(x,y), \\ 
& \dot{y}=-\gamma y+xy-\epsilon y^2, 
\end{split}
\ee 
where: 
\begin{itemize} 
	\item $x$ stands for the liquidity and $y$ stands for the profit, 
	\item $K$, $\gamma$ and $\epsilon$ are positive constants describing, as usually, carrying capacity of the environment $K$, decay rate in profit$\gamma$ and self-competition coefficient in profit $\epsilon$, 
	\item the function $\mcal{L}$ is considered to be non-negative and describing the decrease of the chance to obtain (typically) a bank loan if the liquidity (and profit) falls below some threshold.    
\end{itemize} 

Semmler and Sieveking have used the following kind of the function $\mcal{L}$: 
\bdis 
\mcal{L}(x,y)=L(\max\{0,\alpha x_e-x\}\max\{0,\beta y_e-y\}), 
\edis  
where $(x_e,y_e)$ are the coordinates of the positive-quadrant equilibrium of the model (\ref{SSE}) with $\mcal{L}=0$, $0<\alpha,\beta<1$ are adjustment parameters in the function $\mcal{L}$ and the function $L:\ \mbb{R}^+_0\to\mbb{R}^+_0$ is non-decreasing, continuous function, Semmler and Sieveking have used the following one 
\bdis 
L(t)=v\sqrt{t}, 
\edis  
with positive control parameter $v$. With respect to this one they have discussed the existence of cycles in their model. 

The idea of our work is to replace the non-smooth function $\mcal{L}$ by some smooth function that can effectively serve for the same purpose (decreasing of the chance to gain a bank loan) with some support in established economic considerations. To realize this idea we have decided to consider the Allee effect. This effect has originated in experimental biology, see \cite{Allee}, and has been widely studied in the next about hundred years. The point of the effect is the decrease of the reproduction rate of (single) population dynamics due to low value of the population (density). The effect is applied also in models of dynamics in economics, see for example \cite{LuisEtAl}.  

In the main body of this text, we will use consistently the notions from the standard biological population dynamics (predators, preys) instead of those from macroeconomics (liquidity, profit).

\subsection{Preliminary results} 

\subsubsection{} 

We start with the Lotka-Volterra model affected by the strong Allee effect. Denoting $x$ and $y$ as the populations (densities) of prey and predator, respectively, we have to deal with the dynamical system 
\be \label{i1} 
\begin{split}
& \dot{x}=x-L-xy, \\ 
& \dot{y}=-\gamma y+xy, 
\end{split} 
\ee 
where $\gamma$ and $L$ are positive parameters with the standard meaning: $L$ is the threshold for the population of preys to be reproduced, and $\frac{1}{\gamma}$ is the typical survival time of the predator in absence of preys. This model possesses single positive quadrant equilibrium $E_0$, namely 
\bdis 
E_0\equiv(x_0,y_0)=\left(\gamma,1-\frac{L}{\gamma}\right) 
\edis 
only if $L<\gamma$. Eigenvalues of the Jacobian matrix at $E_0$ read 
\bdis 
\lambda_0^\pm=\frac{1}{2}(p\pm\sqrt{p^2-4\gamma+4p\gamma}), 
\edis  
where 
\bdis 
L=p\gamma,\ 0<p<1, 
\edis  
and this means that the equilibrium $E_0$ of the model (\ref{i1}) is linearly unstable, and therefore no Hopf bifurcation takes place in the model (\ref{i1}). In other words: the strong Allee effect applied on preys in the Lotka-Volterra model turns the neutral Lotka-Volterra equilibrium into the unstable one. 

\subsubsection{} 

Now, let us turn our attention to Lotka-Volterra model modified by the weak Allee effect as follows 
\be \label{i2} 
\begin{split}
& \dot{x}=F(x)x-xy, \\ 
& \dot{y}=-\gamma y+xy, 
\end{split}
\ee 
where 
\bdis 
F:\ \mbb{R}^+_0\to [0,1] 
\edis 
is an increasing differentiable function such, that $F(0)=0$ and $F(x)\xrightarrow{x\to+\infty}1$. The first condition means that the effective reproduction ratio is suppressed to zero at very small populations (densities), and the second one means that at very high populations (densities) the constant reproduction ratio is restored. This model has again an positive quadrant equilibrium 
\bdis 
E_0\equiv(x_0,y_0)=(\gamma,F(\gamma))
\edis 
with the related Jacobian matrix 
\bdis 
J_0=\begin{pmatrix}
	\gamma F'(\gamma) & -\gamma \\ F(\gamma) & 0
\end{pmatrix}. 
\edis 
First of all, the trace of the Jacobian matrix 
\bdis 
\tr J_0=\gamma F'(\gamma) 
\edis  
is always positive and therefore no Hopf bifurcation can occur at $E_0$. Further, both eigenvalues of $J_0$: 
\bdis 
\lambda_0^\pm=\frac{1}{2}\gamma F'(\gamma)\left[1\pm\sqrt{1-4\frac{F(\gamma)}{\gamma F'^2(\gamma)}}\right]
\edis 
have positive real parts. Hence, $E_0$ is linearly unstable. Again, the Allee effect (now the weak one) turns the original neutral Lotka-Volterra equilibrium into the unstable one. 

\subsubsection{} Finally, within these preliminary results, we will investigate properties of the positive quadrant equilibrium of a Lotka-Volterra model with the weak Allee effect in preys and logistic bound on prey population. Our equations read 
\be \label{i3} 
\begin{split} 
& \dot{x}=F(x)x\left(1-\frac{x}{K}\right)-xy, \\ 
& \dot{y}=-\gamma y+xy, 
\end{split} 
\ee 
where $K>0$ is the carrying capacity of the environment. It is very well-known that the model (\ref{i3}), in the case $F(x)=1$, i.e. without the Allee effect, has single single predator-free equilibrium $(K,0)$ that is stable provided there is no positive-quadrant equilibrium, and it becomes unstable when positive quadrant equilibrium appears, see for example \cite{Bazykin}. For the positive quadrant equilibrium $E_0\equiv (x_0,y_0)$ of the system (\ref{i3}) we easily obtain the following conditions 
\bdis 
x_0=\gamma \ \wedge \ y_0=F(\gamma)\left(1-\frac{\gamma}{K}\right), 
\edis 
which require that 
\bdis 
\gamma<K. 
\edis 
The Jacobian matrix of the model (\ref{i3}) at $E_0$ reads 
\bdis 
J_0=\begin{pmatrix}
	F'(\gamma)\gamma(1-\frac{\gamma}{K})-\frac{\gamma F(\gamma)}{K}  & -\gamma \\ F(\gamma)(1-\frac{\gamma}{K}) & 0
\end{pmatrix}. 
\edis 
Subsequently, for the trace of the matrix $J_0$ we obtain the following asympototic formula 
\bdis 
\tr J_0=F'(\gamma)\gamma+\mcal{O}\left(\frac{1}{K}\right),\ K\to+\infty,  
\edis  
which means that the following statement holds true. 

\begin{lemma}
Provided $K$ is big enough (for fixed $\gamma$), at least one eigenvalue of the matrix $J_0$ has positive real part. 
\end{lemma} 

 In that means, the Allee effect can destabilize the original equilibrium for suitable choice of parameters $\gamma$ and $K$.


\section{Weak Allee effect in the Lotka-Volterra model with logistic bound on population of preys} 

\subsection{} 

Let us consider the dynamical system as follows 
\be \label{s2_1} 
\begin{split}
& \dot{x}=\frac{x^2}{x+M}\left(1-\frac{x}{K}\right)-xy, \\ 
& \dot{y}=-pKy+xy, 
\end{split} 
\ee 
where $M>0$ is the control parameter for the Allee effect, $K>0$ is carrying capacity of the environment, and $p$ is bounded between $0$ and $1$ and the value $\gamma=pK$ defines the decay rate of the predator population starving without preys. This range of $p$ ensures that there is a positive quadrant equilibrium 
\bdis 
E_0=(x_0,y_0)=\left(Kp,\frac{Kp(1-p)}{M+Kp}\right). 
\edis 
The Jacobian matrix of the model (\ref{s2_1}) reads at $E_0$: 
\bdis 
J_0=\begin{pmatrix}
	-\frac{Kp[Kp^2+M(-1+2p)]}{(M+Kp)^2} & -Kp \\ \frac{Kp(1-p)}{M+Kp} & 0
\end{pmatrix}. 
\edis  
This matrix is traceless for $M=M_0$, where 
\be \label{s2_1M0}
M_0=\frac{Kp^2}{1-2p}, 
\ee 
and this $M_0$ is positive if and only if 
\bdis 
0<p<\frac 12. 
\edis 
So, for every $K>0$ and every $0<p<\frac 12$ it is reasonable to verify if the Hopf bifurcation takes place at $(E_0,M_0)$. If $\lambda$ stands for eigenvalues of the Jacobian matrix $J_0$, then direct calculations yield 
\bdis 
\left.\frac{{\rm d}\re\lambda}{{\rm d}M}\right|_{M=M_0}=\frac{1}{2Kp}\frac{(1-2p)^3}{(1-p)^2}\equiv \mcal{A}>0, 
\edis  
and 
\bdis 
\im\lambda|_{M=M_0}=\pm\sqrt{Kp(1-2p)}\equiv\pm\omega \not=0. 
\edis 
Subsequently, this means by the Hopf bifurcation theorem, that our system (\ref{s2_1}) is, in the vicinity of the point $(E_0,M_0)$, equivalent with the system described by the polar form bifurcation equation 
\bdis 
\dot{\rho}=\rho[\mcal{A}(M-M_0)+\mcal{B}\rho^2], 
\edis  
where $\mcal{B}$ is the first Lyapunov quantity. In this case, it is possible to obtain $\mcal{B}$ in the closed form, namely 
\bdis 
\mcal{B}=\frac{p(1-2p)^2}{(1-p)^2}(1-4p). 
\edis 
This first Lyapunov quantity is negative for $p>\frac 14$ and it is positive for $p<\frac 14$. 
Subsequently, we obtain the following statement. 

\begin{theorem}
For every $p\in (0,\frac 12)$, except $p=\frac 14$, our system (\ref{s2_1}) undergoes the Hopf bifurcation at $M=M_0$ given by (\ref{s2_1M0}). Moreover, this bifurcation is supercritical for $p>\frac 14$ and it is subcritical for $p<\frac 14$. 
\end{theorem} 

For $p=\frac 14$ the first Lyapunov quantity $\mcal{B}$ equals zero, and there is a possibility, that some degenerate Hopf bifurcation takes place. In the next lowest order of degeneracy it is the Bautin bifurcation that may take place. To check this assumption we need to find the value of the second Lyapunov quantity $\mcal{C}$ at the point 
\be \label{BP1} 
(M_B(p_B),p_B)\equiv \left(M_B(p_B),\frac 14\right)=\left(\frac K8,\frac 14\right). 
\ee 
This can be done in a closed form with the result 
\bdis 
\mcal{C}=\frac{604-2004\sqrt{2}K^{1/2}+210K-126\sqrt{2}K^{3/2}+63K^2}{486K^3}. 
\edis 
The equation $\mcal{C}=0$ has exactly two roots 
\bdis 
K_-\approx 0.047,\ K_+\approx 18.795. 
\edis 
Hence, the next theorem holds true. 

\begin{theorem} \label{thms2_1_2}
For every positive $K$, except $K=K_{\mp}$, our system (\ref{s2_1}) undergoes the Bautin bifurcation at (\ref{BP1}).  
\end{theorem} 

\begin{figure}[h] 
\centering 
\includegraphics[scale=0.45]{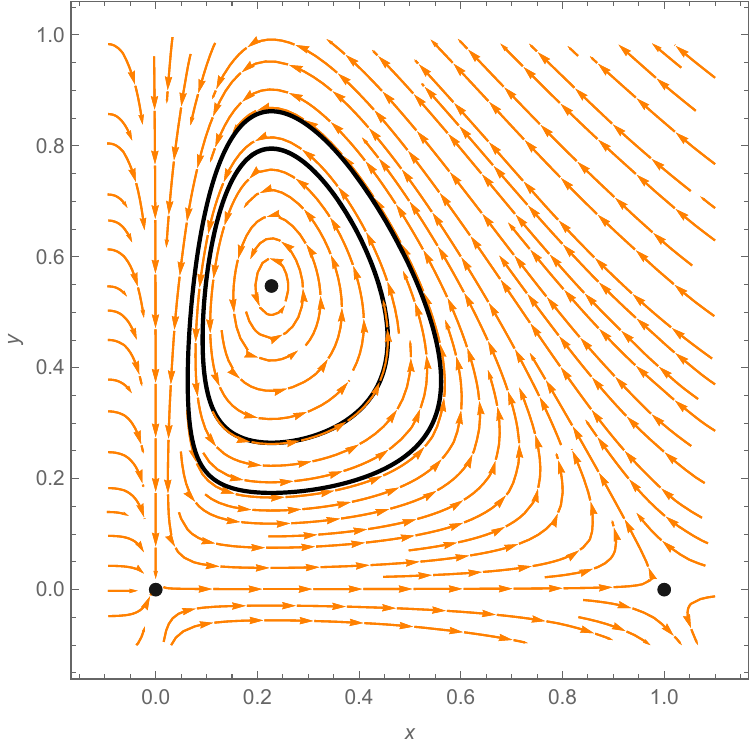} 
\includegraphics[scale=0.45]{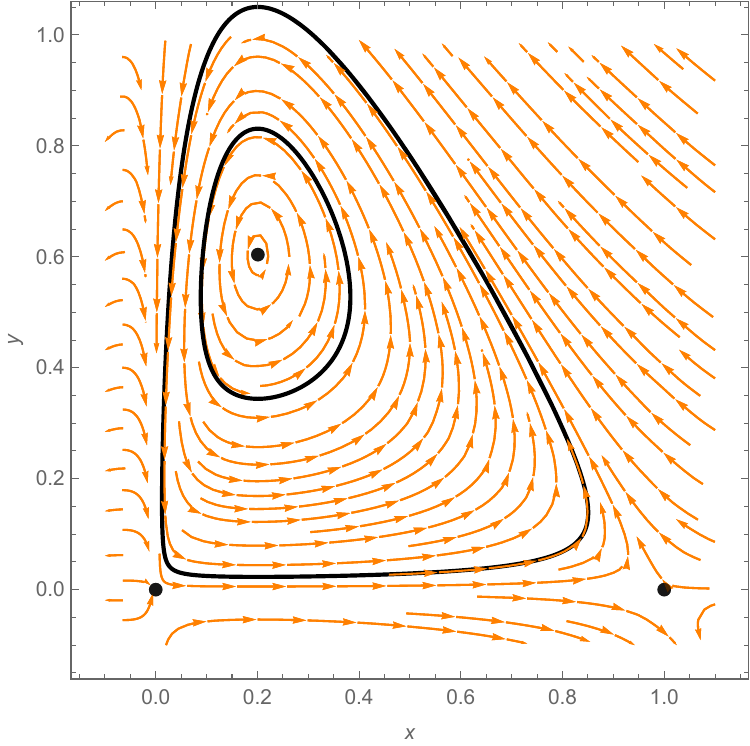} 
\caption{ Bautin cycles of the model (\ref{s2_1}) according to Example \ref{exa1}. } 
\label{f_WA2}
\end{figure}

\begin{example} \label{exa1} 
We have constructed a numerical example to visualize the result of our theorem \ref{thms2_1_2}. Namely, two limit cycles, see Fig. \ref{f_WA2} born via the Bautin bifurcation in the system (\ref{s2_1}) for the following values of the parameters: $K=1$, $M=M_0(1+R\cos U)$, and $p=\frac 14(1+R\sin U)$, $U=0.72\times 2\pi$, and $R=\frac{1}{11}$ for the left panel, and $R=\frac 15$ for the right panel. The equilibrium is locally stable, the inner cycle is unstable and the outer cycle is stable in both cases. 
\end{example} 

\section{Strong Allee effect in the Lotka-Volterra model with logistic bound on population of preys, type I model}  

Now, we shall consider the strong Allee effect in the LV model with the logistic bound on the prey's population 
\be \label{SATI1} 
\begin{split} 
& \dot{x}=(x-L)\left(1-\frac{x}{K}\right)-xy, \\ 
& \dot{y}=-\gamma y+xy, 
\end{split} 
\ee 
where the positive constant $L$ is the control parameter governing the Allee effect. This model has, for $\gamma<K$ the single positive-quadrant equilibrium only for $L<\gamma$. We can use the last two inequalities  for suitable reparametrization of the model as follows: 
\bdis 
\gamma = qK,\ L=p\gamma=pqK,\ p,q\in (0,1). 
\edis 
The positive-quadrant equilibrium $E_0=(x_0,y_0)$ is then given by the following equation  
\be \label{SATI2} 
E_0=\left( Kq,(1-p)(1-q)\right). 
\ee 
Except this equilibrium, our model has two more equilibria that are predator-free, namely 
\bdis 
E_K=(K,0),\ E_L=(L,0)=(pqK,0). 
\edis 
If we denote the eigenvalues of related Jacobian matrices at these equilibria as $\lambda^K$ and $\lambda^L$, respectively, then simple calculations lead to their values 
\bdis 
\lambda^K = \{K(1-q),-1+pq\},\ \lambda^L=\{-K(1-p),1-pq\}. 
\edis  
We can conclude: 
\begin{lemma}
Provided the model (\ref{SATI1}) has the positive-quadrant equilibrium, then the equilibria $E_K$ and $E_L$ are unstable. 
\end{lemma}

The trace of related Jacobian matrix $J_0$ at $E_0$ (\ref{SATI2}) is then 
\bdis 
\tr J_0=p-q, 
\edis  
and it vanishes for $p=p_0=q$. Further, the eigenvalues $\lambda_\pm$ of $J_0$ at the critical value $p=p_0$ are pure imaginary, namely 
\bdis 
\pm i\omega_0,\ \omega_0=\sqrt{Kq}(1-q). 
\edis 
Moreover, the transversality condition is met since 
\bdis 
\left. \frac{{\rm d}}{{\rm d}p}\re\{\lambda_\pm\}\right|_{p=p_0}=\frac 12\not=0. 
\edis 
Subsequently, $p=p_0$ is the Hopf point of our system (\ref{SATI1})  with the locally stable equilibrium $E_0$ for $p<p_0$ and unstable equilibrium for $p>p_0$, and we can check for the type of the Hopf bifurcation at this point calculating the first Lyapunov quantity $\mcal{B}$ with the result 
\bdis 
\mcal{B}=q(1-q^2)>0. 
\edis 
As a consequence, we have the next theorem. 

\begin{theorem}
For every positive $K$ and every $q\in (0,1)$, our system (\ref{SATI1}) undergoes the subcritical Hopf bifurcation at $p=q$ with locally stable equilibrium for $p<q$. 
\end{theorem}

\section{Strong Allee effect in the Lotka-Volterra model with logistic bound on population of preys, type II model} 

The origin $(0,0)$ is not an equilibrium of the model (\ref{SATI1}). This is not necessarily a problem since the the main attention is to be paid to the dynamics in the vicinity of the positive-quadrant equilibrium, that is far-away from the origin. On the other hand, there are ways how to bring the origin back as an equilibrium within a kind of the strong Allee effect in Lotka-Volterra type dynamics. One simple possibility is as follows. We shall consider the model 
\be \label{SATII1} 
\begin{split}
& \dot{x}=\frac{x^2(x-L)}{(x+M)^2}\left(1-\frac{x}{K}\right)-xy, \\ 
& \dot{y}=-\gamma y+xy,
\end{split}
\ee 
with positive constants $K,L,M$ and $\gamma$. The constants $L$ and $M$ are the control parameters of the Allee effect. In what follows we shall assume that 
\bdis 
L<K. 
\edis 
This assumption is very natural, it requests that the Allee effect is in charge below the carrying capacity of the environment. Next, the positive-quadrant equilibrium $E_0$ of the model (\ref{SATII1}) must be given by the formula 
\be \label{SATII2} 
E_0\equiv(x_0,y_0)=\left(\gamma,\frac{-\gamma(\gamma-K)(\gamma-L)}{K(\gamma+M)^2}\right), 
\ee  
and is present for 
\be \label{SATII3} 
L<\gamma<K. 
\ee 
Motivated by the inequalities (\ref{SATII3}) it is suitable to re-parametrize the model (\ref{SATII1}) by means of parameters $(K,p,q)$: 
\be \label{SATII4} 
\gamma=pK,\ L=q\gamma=pqK,\ 0<p,q<1. 
\ee 
The positive-quadrant equilibrium (\ref{SATII2}) then takes the following form  
\be \label{SATII5} 
E_0=\left(pK,p^2K^2\frac{(1-p)(1-q)}{(pK+M)^2}\right). 
\ee 
Except $E_0$ the model (\ref{SATII1}) has three other equilibria, namely 
\bdis 
E_o=(0,0),\ E_K=(K,0),\ E_L=(L,0)=(pqK,0). 
\edis 
The eigenvalues of the related Jacobian matrices are as follows: 
\bdis 
\begin{split} 
& \lambda^o=\{-Kp,0\},\ \lambda^K=\left\{K(1-p),-\frac{K^2}{(K+M)^2}(1-pq)\right\}, \\ 
& \lambda^L=\left\{\frac{K^2p^2q^2}{(M+Kpq)^2}(1-pq),-Kp(1-q)\right\}. 
\end{split} 
\edis 
The equilibria $E_K$ and $E_L$ are therefore linearly unstable, and the equilibrium $E_o$ is a weak attractor since for $x$ positive, and small enough, we obtain 
\bdis 
\dot{x}=-\frac{pqK}{M^2}x^2+\mbox{h.o.t}. 
\edis 
The Jacobian matrix at the positive-quadrant equilibrium (\ref{SATII5}) reads 
\bdis 
J_0=\begin{pmatrix}
-\frac{K^2p^2[Kp(p-q)+M(-2+p(3-2q)+q)]}{(M+Kp)^3} & -Kp \\ 
\frac{K^2p^2(1-p)(1-q)}{(M+Kp)^2} & 0
\end{pmatrix}, 
\edis 
and its trace is equal to zero if and only if 
\be \label{SATII6} 
M=M_0=\frac{Kp(p-q)}{2-q+p(-3+2q)}\equiv K\times \kappa. 
\ee 
Then the matrix 
\bdis 
J_0(M_0)=\begin{pmatrix}
	0 & -Kp \\ 
	\frac{K^2p^2(1-p)(1-q)}{(M+Kp)^2} & 0
\end{pmatrix}, 
\edis 
has a pair of imaginary eigenvalues $\lambda^0=\pm i\omega$ with 
\bdis 
\omega=\frac 12\sqrt{\frac{Kp}{(1-p)(1-q)}}(2-q+p(-3+2q)). 
\edis 
Next, we have to require the positiveness of the critical value $M_0$, see (\ref{SATII6}). This requirement results in the following bounds on parameters $p$ and $q$: 
\be \label{SATII7} 
\kappa>0 \ \Leftrightarrow \ \{ q<p,\ 2-q+p(-3+2q)>0\}. 
\ee 

Now, we are prepared to construct the polar form of Hopf bifurcation equation for our model (\ref{SATII1}) at $E_0$ and $M_0$ 
\bdis 
\dot{\rho}=\rho[\mcal{A}(M-M_0)+\mcal{B}\rho^2]. 
\edis 
There are explicit results for the coefficients $\mcal{A}$ and $\mcal{B}$, namely 
\be \label{SATII8} 
\begin{split}
& \frac{\mcal{A}}{K^2}=p^2(1-p)(1-q)(-2+q+p(3-2q)), \\ 
& \mcal{B}=\frac{(-2+q+p(3-2q))^4}{64(1-p)^5(1-q)^5}\times \beta , \\ 
& \beta=-[q^2(q-4)+p^2(-16+29q-4q^2)+ \\ 
& p^3(9-16q+4q^2)+p(4-4q-13q^2+4q^3)]. 
\end{split}
\ee 
It is easy to see that for positive $M_0$, see (\ref{SATII7}), the coefficient $\mcal{A}$ is also positive. The sign of the coefficient $\mcal{B}$, the first Lyapunov quantity, is the same as the sign of $\beta$.  

\begin{figure}[h] 
	\centering 
	\includegraphics[scale=0.55]{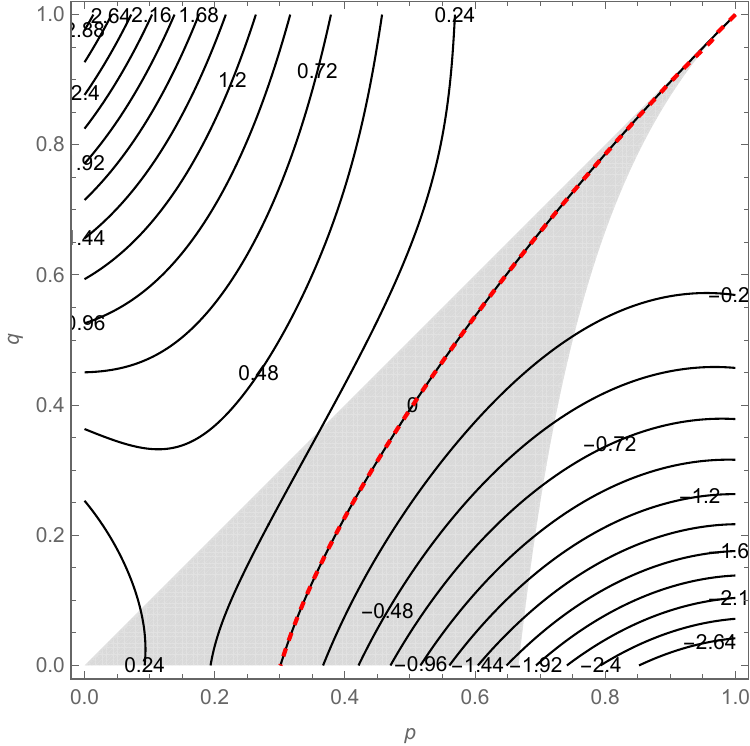} 
	\caption{Region of positiveness of the bifurcation value $M_0$, (\ref{SATII6}) and level lines of $\beta$, (\ref{SATII8}). The first Lyapunov quantity changes its sign on the curve $\beta=0$ (shown as red dashed line). } 
	\label{f_SA2} 
\end{figure} 

Consequently, we have shown that the following theorem holds true. 

\begin{theorem}
For every positive $K$  our system (\ref{SATI1}) undergoes the Hopf bifurcation at $E_0$ and $M=M_0$. $E_0$ is stable for $M<M_0$ and the bifurcation is supercritical for the pairs $(p,q)$ lying below the curve $\beta=0$, see Fig. \ref{f_SA2}, and it is subcritical above the curve $\beta=0$. 
\end{theorem} 

\begin{figure}[h] 
	\centering 
	\includegraphics[scale=0.45]{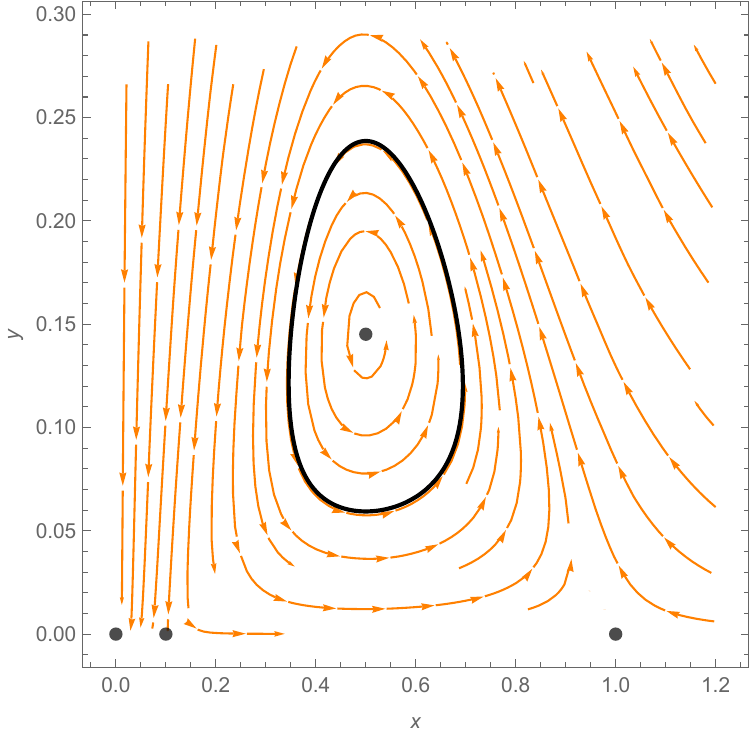} 
	\includegraphics[scale=0.45]{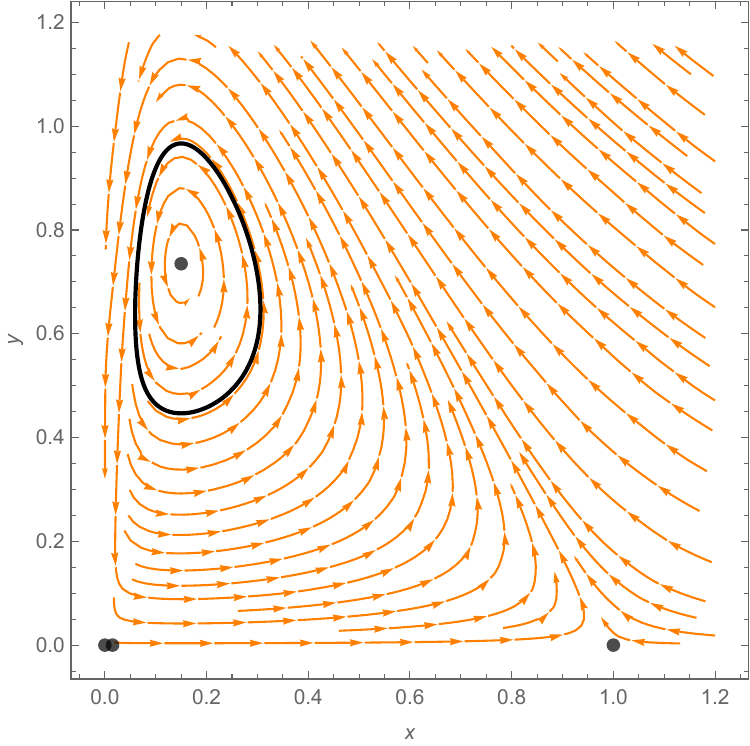}
	\caption{Stable (left panel) and unstable (right panel) cycles of the model (\ref{SATII1}) according to Example \ref{exa2}.} 
	\label{f_SA3} 
\end{figure}

\begin{example} \label{exa2} 
According to formulae (\ref{SATII8}), we can construct examples of dynamics of the system (\ref{SATII1}) containing the stable cycle and also the unstable cycle. We can take $K=1$  and then: 
\begin{itemize}
	\item we take $p=0.2$ and $q=0.2$ and appropriate value of $M$, namely $M=1.1\times M_0$ to obtain the stable Hopf cycle, see the left panel of the Fig. \ref{f_SA3}; 
	\item we take $p=0.15$ and $q=0.1$ and appropriate value of $M$, namely $M=0.6\times M_0$ to obtain the unstable Hopf cycle, see the right panel of the Fig. \ref{f_SA3}
\end{itemize}
\end{example}

\section{Leslie-Holling type model with the weak Allee effect}

Now, let us consider the Leslie-Holling type model with the weak Allee effect described by the system of equations 
\be \label{LHWA1} 
\begin{split}
	& \dot{x}=\frac{x^2}{x+M}-xy, \\ 
	& \dot{y}=ry\left(1-\frac{y}{x}\right),
\end{split}
\ee 
where $M>0$ is the control parameter for the Allee effect and $r$ is the effective rate of reproduction of the predator, modulated by the logistic-like term $1-\frac{y}{x}$, where the population of the prey serves as the carrying capacity for the population of  predator. The model (\ref{LHWA1}) has one positive-quadrant equilibrium for every positive $r$ and every $0<M<1$, namely 
\be \label{LHWA2} 
E_0=(x_0,y_0)=(1-M,1-M). 
\ee 
Otherwise, $M\geq 1$, there is no positive-quadrant equilibrium in the model (\ref{LHWA1}). The Jacobian matrix of the model (\ref{LHWA1}) at $E_0$ takes the form 
\bdis 
J_0=\begin{pmatrix}
	(1-M)M & -1+M \\ r & -r 
\end{pmatrix},  
\edis 
and its trace is given by the formula 
\bdis 
\tr J_0=(1-M)M-r. 
\edis 
The matrix is traceless if and only if 
\bdis 
r=r_0=(1-M)M, 
\edis  
and the eigenvalues of the matrix $J_0$ at $r=r_0$ are purely imaginary given as 
\bdis 
\lambda_\pm=\pm i \omega,\ \omega=\sqrt{M(1-M)^3}. 
\edis 
Evaluation of the coefficients in the polar form of the Hopf bifurcation equation is quite straightforward in this case and results in 
\be \label{LHWA3} 
\dot{\rho}=\rho\left\{-\frac 12(r-r_0)-\frac{1}{2}M^3(1-M)^4\rho^2\right\}, 
\ee  
which allows us to formulate the next theorem. 

\begin{theorem}
Provided $0<M<1$ the system (\ref{LHWA1}) undergoes non-degenerate and supercritical Hopf bifurcation at the positive-quadrant equilibrium $E_0$ and at $r=r_0=M(1-M)$ with the stable equilibrium for $r>r_0$. 
\end{theorem} 

\begin{figure}[h] 
	\centering 
	\includegraphics[scale=0.55]{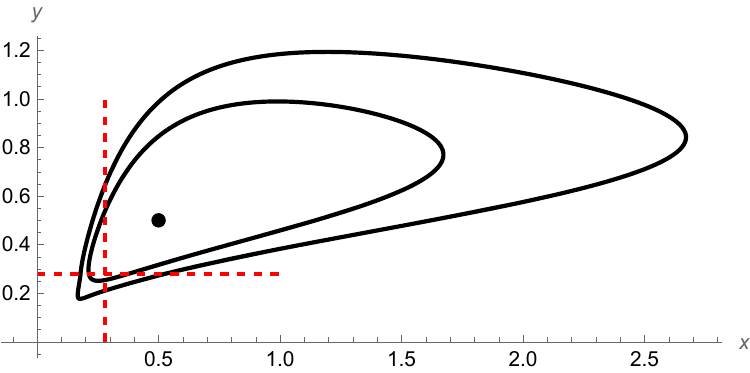} 
	\caption{Red dashed lines are the boundaries of the support of the function $\mcal{L}(x,y)=0.6\times [\max\{0,\tilde{x}-x\}\max\{0,\tilde{y}-y\}]^{1/2}$, where $\tilde{x}=\tilde{y}=\frac{x_0}{1.8}$. Next, $r=0.94\times r_0$ and $M=\frac 12$. Both cycles are stable. The inner one is the Hopf cycle of the system (\ref{LHWA1}) and the outer cycle appears in the system (\ref{LHWA4}).} 
	\label{f_LHWA1} 
\end{figure}

\begin{example}
We provide the reader with an example of the dynamics of the system (\ref{LHWA1}) with the stable Hopf cycle and the next system 
\be \label{LHWA4} 
\begin{split}
& \dot{x}=\frac{x}{M+x}x-xy-x\mcal{L}(x,y), \\ 
& \dot{y}=ry\left(1-\frac{y}{x}\right), 
\end{split}
\ee 
where $\mcal{L}$ stands for the reactive term of Semmler and Sieveking, \cite{SemmlerSieveking}. In effect, if the stable Hopf cycle collides with the support of the Semmler - Sieveking reactive term, it significantly blows keeping its stability. The situation is described on the Fig. \ref{f_LHWA1}. 
\end{example}

\section{Leslie-Holling type model with the strong Allee effect}

\begin{figure}[h] 
	\centering 
	\includegraphics[scale=0.55]{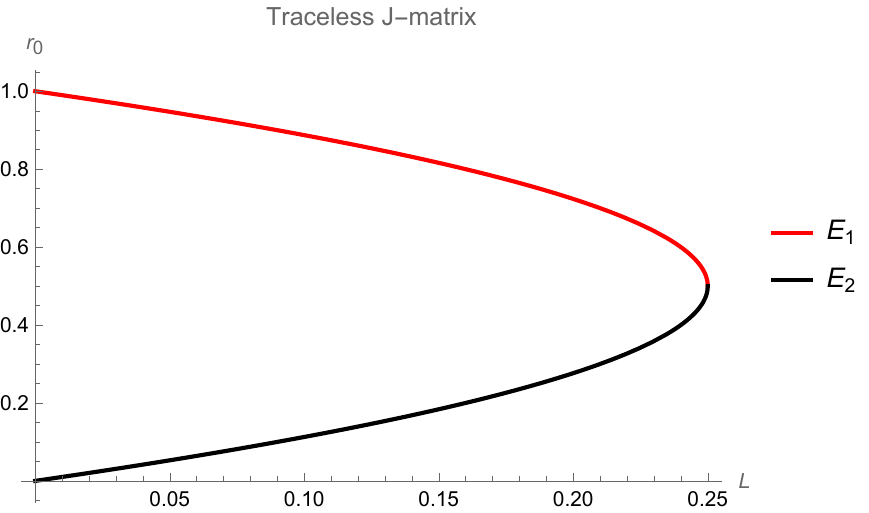} 
	\caption{Curves of vanishing traces of the Jacobian matrices $J_1$ (red) and $J_2$ (black) in the parametric space $(r,L)$ according to the formulae (\ref{LHSA3}).} 
	\label{f_LHSA1} 
\end{figure}

In this section we consider the following model 
\be \label{LHSA1} 
\begin{split}
& \dot{x}=x-L-xy, \\ 
& \dot{y}=ry\left(1-\frac{y}{x}\right), 
\end{split}
\ee 
with the control parameter $L>0$ of the Allee effect. Provided $0<L<\frac 14$ this model has two distinct positive-quadrant equilibria 
\be \label{LHSA2} 
\begin{split}
& E_1=(x_1,y_1)=\left(\frac{1-\sqrt{1-4L}}{2},\frac{1-\sqrt{1-4L}}{2}\right), \\ 
& E_2=(x_2,y_2)=\left(\frac{1+\sqrt{1-4L}}{2},\frac{1+\sqrt{1-4L}}{2}\right). 
\end{split}
\ee  
These equilibria collide at $L=\frac 14$ and further, for $L>\frac 14$, the model (\ref{LHSA1}) has no positive-quadrant equilibrium. Jacobian matrices related to the equilibria (\ref{LHSA2}) are as follows 
\bdis 
\begin{split} 
& J_1=\begin{pmatrix}
	\frac 12(1+\sqrt{1-4L}) & \frac 12(-1+\sqrt{1-4L}) \\ r & -r
\end{pmatrix}, \\ 
& J_2=\begin{pmatrix}
	\frac 12(1-\sqrt{1-4L}) & \frac 12(-1-\sqrt{1-4L}) \\ r & -r
\end{pmatrix}. 
\end{split} 
\edis 
These matrices are traceless for $r=r_{01}$ and $r=r_{02}$, respectively, where 
\be \label{LHSA3} 
r_{01}=\frac{1}{2}(1+\sqrt{1-4L}),\ r_{02}=\frac{1}{2}(1-\sqrt{1-4L}), 
\ee  
see the Fig. \ref{f_LHSA1}. 

\begin{figure}[h] 
	\centering 
	\includegraphics[scale=0.45]{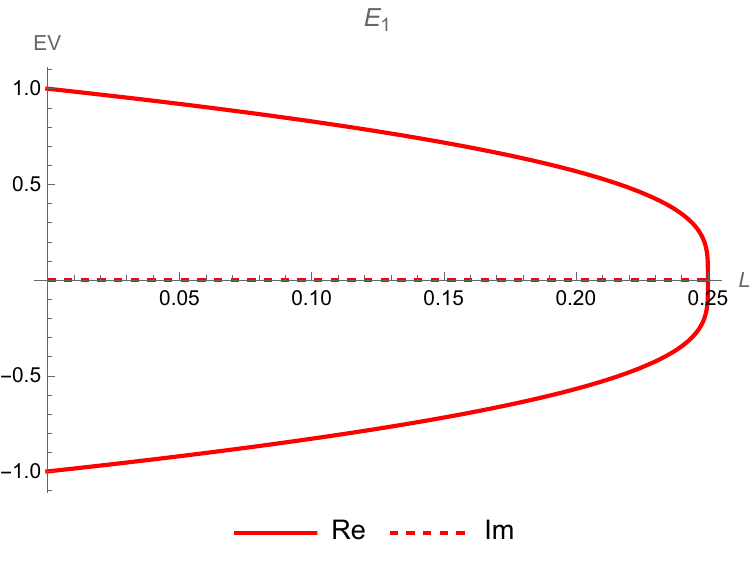} 
	\includegraphics[scale=0.45]{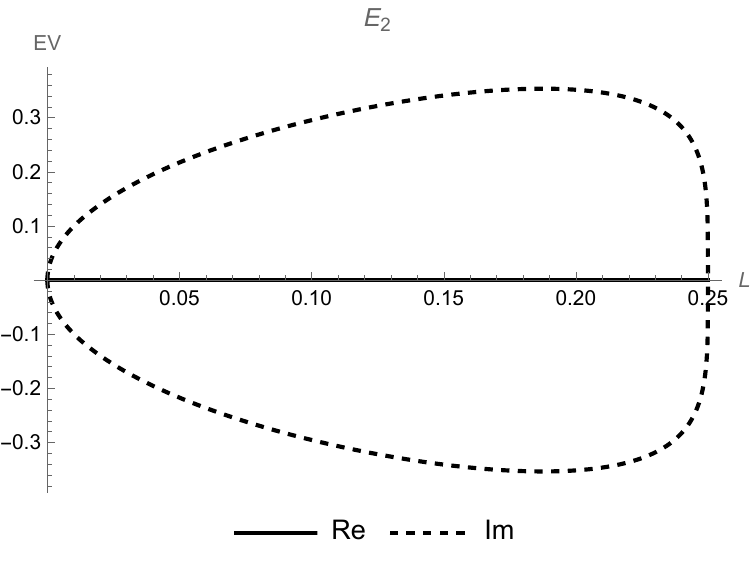} 
	\caption{Eigenvalues of the traceless versions of the matrices $J_1$ and $J_2$, respectively.} 
	\label{f_LHSA2} 
\end{figure}

Eigenvalues of the matrices $J_1(r=r_{01})$ and $J_2(r=r_{02})$ are given as follows 
\be \label{LHSA4} 
\begin{split}
& \lambda_\pm^{(1)}=\pm \Omega,\ \Omega=\frac{\sqrt{1+\sqrt{1-4L}-4L}}{\sqrt{2}}, \\ 
& \lambda_\pm^{(2)}=\pm i\omega,\ \omega=\frac{\sqrt{\sqrt{1-4L}+4L-1}}{\sqrt{2}}. 
\end{split}
\ee 
The result is such that (see also Fig. \ref{f_LHSA2}) the values $\lambda^{(1)}$ are real and the values $\lambda^{(2)}$ are imaginary. As a consequence we can state that no Hopf bifurcation occurs at the equilibrium $E_1$ and the Hopf bifurcation is possible at the equilibrium $E_2$. 

To clarify the question of Hopf bifurcation at the equilibrium $E_2$, we again construct the polar form of the Hopf bifurcation equation with the following result 
\be \label{LHSA5} 
\begin{split}
& \dot{\rho}=\rho\left\{-\frac{1}{2}(r-r_{02})+\mcal{B}\rho^2\right\}, \\ 
& r_{02}=\frac{1}{2}(1-\sqrt{1-4L}), \\ 
& \mcal{B}=\frac{(1-\sqrt{1-4L})(1-2L-\sqrt{1-4L})}{2(1+\sqrt{1-4L})}>0,  
\end{split}
\ee 
which allows us to formulate the next theorem. 

\begin{theorem}
Provided $0<L<\frac 14$, our system (\ref{LHSA1}) undergoes the subcritical Hopf bifurcation at the equilibrium $E_2$, (\ref{LHSA2}), and at the critical value $r=r_{02}$, (\ref{LHSA3}), with stable equilibrium for $r>r_{02}$. 
\end{theorem} 

\section*{Conlusion} 

In this paper we have reviewed two dimensional predator-prey continuous models with the Allee effect in its weak and strong meaning to propose alternatives or additions to Semmler-Sieveking \cite{SemmlerSieveking} model of aggregate-type dynamics of liquidity and profit. According to the paper of Semmler and Sieveking the main attention was paid to what is happening in the vicinity of the positive-quadrant equilibrium by means of the Hopf bifurcation theorem. We have obtained a variety of explicit results about the Hopf bifurcation, non-degenerate as well as degenerate (Bautin) in the considered models.

\noindent 
\thanks

The paper was financially supported by the grant scheme of the Slovak Ministry of Education, project VEGA no. 1/0084/23.

\end{document}